\documentclass[prl, twocolumn, preprintnumbers,amsmath,showpacs,amssymb,superscriptaddress]{revtex4-2}
\usepackage{epsfig}
\usepackage{color}
\usepackage{graphicx}% Include figure files
\usepackage{dcolumn}% Align table columns on decimal point
\usepackage{bm}% bold math
\usepackage{tabularx}% bold math
\usepackage[pdftex, colorlinks, citecolor=blue]{hyperref}   % this is for pdflatex pdf format
\usepackage{appendix}
\usepackage{diagbox}
\usepackage{threeparttable}
\usepackage{multirow}
\usepackage{tabularx}
\usepackage{booktabs}
\usepackage{txfonts}
\usepackage{xcolor}
\usepackage{amssymb}
\usepackage{amsmath}
\usepackage{latexsym}
\usepackage{amsbsy}
\usepackage{array}
\usepackage{esvect} %vector
\usepackage{extarrows}
\usepackage{float}

\begin{document}

%\title{Ferroelectric Switchable Altermagnetism in monolayer CrPS$_3$}%
\title{Two-dimensional Dual-Switchable Ferroelectric Altermagnets: Altering Electrons and Magnons}%

\author{ShuaiYu Wang}
\email{These authors contributed equally to this work.}
\affiliation{Research Center for Quantum Physics and Technologies, School of Physical Science and Technology, Inner Mongolia University, Hohhot 010021, China}

\author{Wei-Wei Wang}
\email{These authors contributed equally to this work.}
\affiliation{Research Center for Quantum Physics and Technologies, School of Physical Science and Technology, Inner Mongolia University, Hohhot 010021, China}

\author{Jiaxuan Fan}
\affiliation{Research Center for Quantum Physics and Technologies, School of Physical Science and Technology, Inner Mongolia University, Hohhot 010021, China}

\author{Xiaodong Zhou}
\affiliation{School of Physical Science and Technology, Tiangong University, Tianjin 300387, China}

\author{Xiao-Ping Li}
\email{xpli@imu.edu.cn}
\affiliation{Research Center for Quantum Physics and Technologies, School of Physical Science and Technology, Inner Mongolia University, Hohhot 010021, China}
\affiliation{Key Laboratory of Semiconductor Photovoltaic Technology and Energy Materials at Universities of Inner Mongolia Autonomous Region, Inner Mongolia University, Hohhot 010021, China}

\author{Lei Wang}
\email{lwang@imu.edu.cn}
\affiliation{Research Center for Quantum Physics and Technologies, School of Physical Science and Technology, Inner Mongolia University, Hohhot 010021, China}
\affiliation{Inner Mongolia Key Laboratory of Microscale Physics and Atom Innovation, Inner Mongolia University, Hohhot 010021, China}

\begin{abstract}
Ferroelectric altermagnets (FEAMs) offer unique magnetoelectric coupling properties by combining the characteristics of both antiferromagnets and ferromagnets, yet their multifunctional electric control remains largely unexplored. Here, we introduce and investigate a scenario for the simultaneous electrical switching of electronic spin and magnonic chirality splitting in two-dimensional FEAMs. Based on the C2DB database, employing symmetry analysis and first-principles calculations, we study prototypical candidates CrPS$_3$ and V$_2$I$_2$O$_2$BrCl. We identify the mechanism: ferroelectricity arises from asymmetric displacements (P along $z$ in CrPS$_3$, V along the $xy$-direction in V$_2$I$_2$O$_2$BrCl), which inherently couples electric polarization to both electronic and magnonic degrees of freedom by retaining [C$_2$$||$M] symmetry. Our calculations explicitly demonstrate that reversing the ferroelectric polarization concurrently switches the sign of the electronic spin splitting and chirality of magnonic modes. This shows these materials as dual-switchable FEAMs, enabling unified electrical manipulation of electron and magnon properties. A potentially experimentally detectable method via the magneto-optical Kerr effect was derived. This work provides a materials-specific realization and theoretical basis for designing novel electrically controlled multifunctional spintronic, spin caloritronic, and magnonic devices.
\end{abstract}

\maketitle

\textit{Introduction} --- Multiferroic materials, uniting ferroelectricity and magnetism (ferromagnetism or antiferromagnetism), promise unprecedented electrical control over magnetic states~\cite{multiferro-1994,eerenstein_multiferroic_2006,multiferroic_2015}. While ferroelectric (FE) antiferromagnetic (AFM) are well-studied, achieving robust coupling between ferroelectricity and antiferromagnetism remains elusive. Altermagnets (AMs), a new collinear antiferromagnetic class\cite{AM-PRX-2022, AM-PRX2-2022, krempasky_altermagnetic_2024, zhou_manipulation_2025, bai_altermagnetism_2024, liu_inverse_2024, zhang_simultaneous_2024, guo_direct_2024, song_altermagnets_2025, jiang_metallic_2025, zhang_crystal-symmetry-paired_2025, ABAM_PRL_RuO2, qian_fragile_2025, bl_AM_PRL, tw_AM_PRL}, in which opposite spin atoms are spatially connected by rotation or mirror symmetry instead of inversion or transition in conventional AFM, beyond AFM and ferromagnets (FM), offer a compelling alternative by intrinsically allowing coexistence of electric polarization and local spin polarization\cite{sun_TypeII-FEAM}, paving the way for efficient gate-controlled spintronics. Symmetry dictates that integrating polar and spin space groups is key, with examples found in perovskites and Ruddlesden-Popper phases\cite{BR-popper_2024}, and theoretical predictions pointing towards switchable bulk FEAMs~\cite{FESAM-Liu}. Extending multiferroicity and altermagnetism to the two-dimensional (2D) limit, potentially via van der Waals heterostructures\cite{sun_TypeII-FEAM} or monolayer\cite{Duan_AFMAM_2025,zhu_FEAM_2025,wang_electric-field-induced_2025}, is crucial for nanoscale device integration. However, realizing intrinsic FE altermagnetism within a single 2D monolayer remains a significant challenge.

On the other hand, the magnonic landscape of AMs is equally rich, offering distinct advantages over conventional magnetic systems\cite{SSE-SNE-magnon1, SSE-SNE-magnon2}. Theoretical models predict~\cite{magnon_RuO2, sodequist_two-dimensional_2024, wang_alternating_2024} that these materials host chiral magnons capable of ultrafast propagation (akin to AFM)~\cite{magnon_RuO2}. Furthermore, the specific symmetries broken in AMs intrinsically lift the degeneracy of left- and right-handed modes, enabling directional spin transport reminiscent of FM\cite{SSE-SNE-magnon1, SSE-SNE-magnon2} $-$ a feature recently verified even at zero external field in certain bulk candidates via techniques like inelastic neutron scattering\cite{INS-MnTe-magnon}. Crucially, theoretical studies highlight that the characteristic spin-split electronic bands inherent to AMs strongly suppress Landau damping~\cite{magnon_RuO2}. This points towards intrinsically low-loss, anisotropic magnon propagation~\cite{costa_giant_2024}, with anisotropy determined by the crystal structure, vital for efficient future magnonic circuits. While foundational concepts like chirality splitting magnonic modes have been experimentally demonstrated in bulk AMs \cite{SSE-magnon, kimel_optical_2024, INS-MnTe-magnon, gray_time-resolved_2024, hortensius_coherent_2021, weber_all_2024} (e.g., through optical techniques and thermal transport measurements), translating this potential to the 2D limit faces significant hurdles. Specifically, reports on chiral magnons in 2D AMs remain scarce. Furthermore, the concept of integrating ferroelectricity for magnonic control in 2D AMs remains unexplored. Such integration promises a powerful paradigm: using FE polarization to concurrently gate both the spin-polarized electrons and the chiral magnons, enabling unprecedented dual control within a single monolayer platform.

\begin{figure}
\centering
\includegraphics[width=0.48\textwidth]{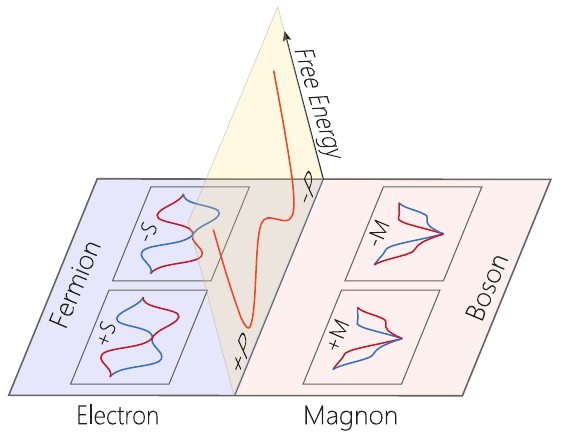}
\caption{Schematic diagram of 2D dual-switchable FEAMs. The FE polarization ($\mathbf{P}$) acts as a master switch, simultaneously controlling both the electronic spin splitting ($\mathbf{S}$) characteristic of altermagnetism and the chiral magnon modes ($\mathbf{M}$). Reversing $\mathbf{P} \to -\mathbf{P}$ forces a corresponding reversal $\mathbf{S} \to -\mathbf{S}$ and $\mathbf{M} \to -\mathbf{M}$, where \textbf{S} and \textbf{M} are the band splitting for the spin of the electron and chirality of the magnon. 
}\label{fig1}
\end{figure} 

In this Letter, we reported the prediction of $2D$ $dual-switchable$ $ferroelectric$ $altermagnets$, where ferroelectricity serves as a master switch for both electronic and magnonic properties intrinsic to altermagnetism. The central finding, depicted in Fig. \ref{fig1}, is that the FE polarization state deterministically dictates not only the sign of the electronic spin splitting but also magnonic chiral mode splitting. This simultaneous electrical control over disparate spin-related phenomena enables nonvolatile manipulation, holding promise for integrated memory architectures. Through combined symmetry analysis and first-principles calculations, we proposed 5 candidates as concrete realizations, and two of them (CrPS$_3$ and V$_2$I$_2$O$_2$BrCl) to explain our strategy. We elucidated how the FE distortion, arising from P and V atom displacements, enforces the symmetry breaking required for the emergence of altermagnetism. The reversibility of this dual electric control is computationally verified, and we suggest magneto-optical Kerr effect and spin Seebeck effect measurements for experimental validation. Our study unveils a fundamental mechanism for electrically controlling altermagnetism and its excitations in 2D, opening avenues for novel spintronic functionalities. The detailed description of the computational methods refers to the Supplemental Material \cite{SM} (see also Refs. \citenum{DFPT1, DFPT2, PHONOPY, VASP, PAW, PBE, VASPKIT, Wannier90, TB2J, PASP, Toth_2015, COLPA1978327, DFTPU} therein).

\textit{Material design principles} --- 
Identifying candidate materials for 2D dual-switchable FEAMs requires establishing the conditions for coexisting ferroelectricity and altermagnetism. Unlike conventional FM or AFM described by magnetic space groups, AMs represent a distinct magnetic class characterized by non-relativistic spin splitting, necessitating a spin space group (SSG) description~\cite{SSG1, SSG2, SSG3}. Furthermore, the FE polarization \textbf{P} and spin polarization \textbf{S}, manifested in the spin splitting $\Delta$E$_k^{S}$ = E$_{\uparrow}$(${k}$) $-$ E$_{\downarrow}$($k$), where subscript $\uparrow$ and $\downarrow$ denote spin up and spin down states, should be coupled and switched simultaneously in FE transition. This means that \textbf{P}$-$\textbf{S} coupling can be described by the ability to synchronously reverse both \textbf{P} ($\mathbf{P} \to -\mathbf{P}$) and \textbf{S} ($\mathbf{S} \to -\mathbf{S}$) via the combined $\mathcal{PT}$ operation, typically connecting two FE polarized states. To address this, we initially focus on the fundamental symmetry requirements for each order parameter independently. Within the SSG framework, three criteria emerge: (i) To exhibit altermagnetism, the material must break both the combined parity-time reversal ($\mathcal{PT}$) symmetry and the time-reversal operations involving fractional translations ($\mathcal{T\tau}$); (ii) For ferroelectricity, the underlying spatial symmetries must belong to a polar space group, allowing for a spontaneous electric polarization \textbf{P}; (iii) For switchability, the kinetic barrier for FE switching must be reasonably low (\textless 0.5 eV) for practical applications. 

Starting from the C2DB database (16789 entries)\cite{C2DB}, we implemented a computational screening [Supplemental Figure S1\cite{SM}] first applying symmetry rules (i) and (ii) outlined above, simplifying the search by initially considering only materials with two magnetic atoms per unit cell. This yielded 12 candidates exhibiting both polar and spin symmetry consistent with altermagnetism [Table SI]. We then addressed the crucial requirements of switchable ferroelectricity and \textbf{P}-\textbf{S} coupling. Using the PSEUDO package\cite{PSEUDO} to analyze potential paraelectric (PE) switching pathways for these 12, we identified 4 candidates where switching involves complex atomic rearrangements (\textgreater 4 atoms), indicating potentially high barriers; these were thus excluded. For the remaining 8 candidates, we rigorously investigated the intrinsic coupling between the FE polarization \textbf{P} and the altermagnetic state \textbf{S}. Remarkably, all 8 investigated candidates exhibited this \textbf{P}-\textbf{S} coupling [Figure S2]. Finally, the climbing image nudged elastic band (CINEB) calculations\cite{CINEB} were performed on these 8 coupled systems, revealing switching barriers exceeding 0.5 eV in 3 candidates, deeming them less practical. Without loss of generality, the classification into three types, based on whether FE displacement involves magnetic atoms (e.g., V$_2$I$_2$O$_2$BrCl), or its first neighbors (e.g., Pt$_2$I$_5$Cl), or second neighbors (e.g., CrPS$_3$), is presented for the 8 confirmed polar AMs before the final kinetic barrier filtering [Table S1].

\textit{Electrical Control of spin splitting in AMs} --- 
\begin{figure*}
\centering
\includegraphics[width=0.95\textwidth]{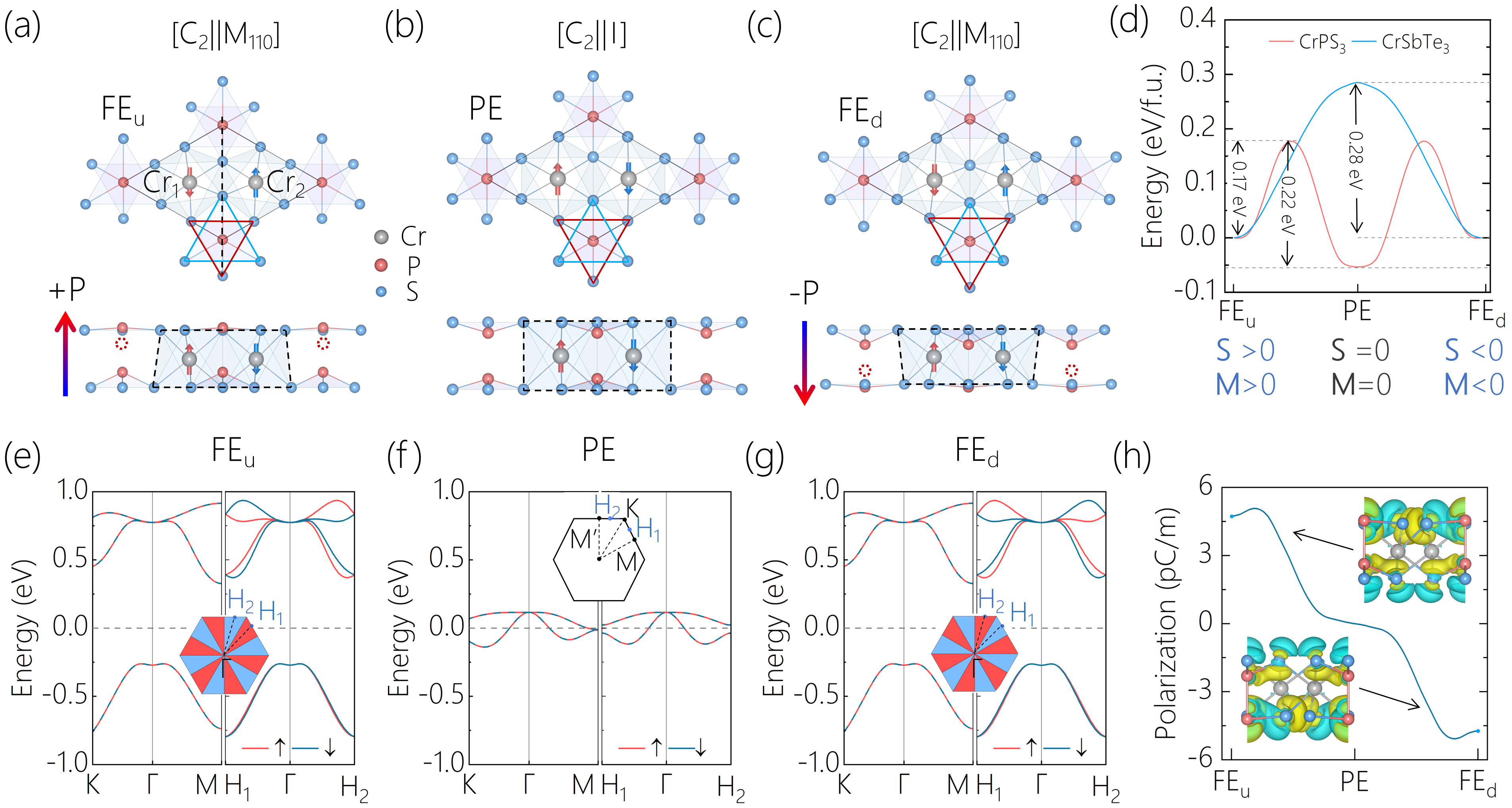}
\caption
{FEAM in CrPS$_3$ monolayer. (a) FE$_u$ and (c) FE$_d$ phases of CrPS$_3$ with a positive and negative polarization, respectively, characterized by a spin group symmetry of [C$_2$$||$M$_{110}$] for two Cr atoms. (b) PE phase is described by spin group symmetry of [C$_2$$||$I] for two Cr atoms. (d) Calculated kinetic switching pathways between FE$_u$ and FE$_d$ states through PE states in CrPS$_3$ and CrSbTe$_3$ under effective Hubbard U$_{eff}$ = 3 eV, where \textbf{S} and \textbf{M} denote the electronic and magnonic band splitting. (e-g) Calculated band structures for (a-c), where the red and blue lines denote the spin-up and -down bands, respectively. The inset in (e) and (g) illustrates the spin-splitting bands in the BZ, indicating an $i$-wave spin-splitting in CrPS$_3$. (h) The FE polarization \textbf{P} of CrPS$_3$ at the kinetic switching pathways in (d), the insert is the charge density differential for the FE$_u$ and FE$_d$ phases, referring to the PE phase, suggesting its polarization. 
}\label{fig2}
\end{figure*}
Next, we detail the coupling mechanism in the identified 2D FEAMs, focusing on CrPS$_3$ as a representative example involving second-neighbor displacement [Fig.~\ref{fig2} (a-c)]. The PE phase of CrPS$_3$ adopts the P$\bar{3}$m1 space group, featuring a centrosymmetric P-P dimer equidistant from the two magnetic Cr atoms. Crucially, the P atoms (second nearest neighbors to Cr) share coordinating S atoms with Cr, providing a structural link for potential magnetoelectric coupling [Fig. \ref{fig2}(a-c)]. The FE phase emerges via a pseudo Jahn-Teller distortion, breaking the [C$_2$$||$$\mathcal{I}$] symmetry: one P atom displaces along the $z$-axis relative to the other, enlarging the upward S3 triangle (cyan)[Fig.~\ref{fig2}(a)], leading to the polar P3m1 space group. This distortion, driven by hybridization between empty Cr-$d$ and occupied S-$p$ states, modifies the Cr coordination environment. Importantly, the loss of inversion symmetry $\mathcal{I}$ between two Cr sublattices in the FE phase lifts the constraints in the PE phase, enabling the altermagnetic state ($\mathbf{S}$) characterized by spin splitting. This contrasts with the experimentally synthesized MnPSe$_3$ monolayer, which remains stable in the centrosymmetric PE phase~\cite{Ni2021_MnPSe3}. To assess the switchability and coupling, we calculated the energy barrier for reversing the polarization ($\mathbf{P}$) between the two degenerate FE states (FE$_u$ $\leftrightarrow$ FE$_d$) using CINEB [Fig.~\ref{fig2}(d)]. Considering pathways via intermediate PE structures, we find a viable switching path with a barrier of $\sim$0.17 eV per formula unit, comparable to other 2D ferroelectrics and indicating robust switchable ferroelectricity. And the PE phase is energetically preferred over the FE phase by 50 meV to form a three-potential well, originating from a magnetic phase transition [Fig.~\ref{fig2}(f) and Figure S5]. And such a magnetic phase transition is material-dependent and does not happen in another candidate CrSbTe$_3$ [Fig.~\ref{fig2}(d)]. 

We now examine the electronic structure evolution in CrPS$_3$. In the PE phase, the crystal structure possesses both inversion ($\mathcal{I}$) and mirror (M$_{110}$) symmetries relating the two Cr sublattices. These symmetries enforce spin degeneracy throughout the Brillouin zone (BZ), ensuring $E_{\uparrow}(\mathbf{k}) = E_{\downarrow}(\mathbf{k})$ for all wavevectors $\mathbf{k}$. Transitioning to the FE phase breaks the [C$_2||\mathcal{I}$] symmetry connecting two Cr sites, while preserving the [C$_2||$M$_{110}$] symmetry. This specific symmetry breaking lifts the constraints imposed by $\mathcal{I}$ symmetry, fulfilling the condition for the emergence of altermagnetism. Consequently, spin splitting appears, i.e., $E_{\uparrow}(\mathbf{k}) \neq E_{\downarrow}(\mathbf{k})$, except possibly along high-symmetry lines protected by the remaining M$_{110}$ symmetry. The calculated band structure [Fig.~\ref{fig2}(e)] indeed exhibits the characteristic momentum-dependent spin splitting along the H1-$\Gamma$-H2 line. Crucially, FE switching between the two degenerate states (FE$_u$ $\leftrightarrow$ FE$_d$) reverses the polarization $\mathbf{P}$ [Fig. \ref{fig2}(h)]. Due to the intrinsic coupling, this reversal simultaneously inverts the sign of the spin splitting: $\Delta E_{FE_u}^{S}(\mathbf{k}) = - \Delta E_{FE_d}^{S}(\mathbf{k})$ [Fig.~\ref{fig2}(g)]. This electrically controlled reversal of spin splitting unequivocally demonstrates robust magnetoelectric coupling between the altermagnetic electronic structure and the FE order in monolayer CrPS$_3$.

In contrast to CrPS$_3$, where out-of-plane (OP) polarization couples to altermagnetism, V$_2$I$_2$O$_2$BrCl exhibits a more complex scenario involving both in-plane (IP) and OP polarization components, corresponding to magnetic atoms themselves and first nearest neighbor, respectively. A switchable IP polarization ($\mathbf{P}_{\mathrm{IP}}$) originates from a pseudo Jahn-Teller distortion affecting the two V sublattices [Figure S6]. Meanwhile, a robust, non-switchable OP polarization ($\mathbf{P}_{\mathrm{OP}}$) arises due to the intrinsic structural asymmetry imposed by the Janus Br-V-Cl bonding along the $z$-axis. The combination of $\mathbf{P}_{\mathrm{IP}}$ and $\mathbf{P}_{\mathrm{OP}}$ results in a low overall symmetry compatible with AM; specifically, the symmetry of the IP FE state permits altermagnetism. Intriguingly, the calculated switching pathway for $\mathbf{P}_{\mathrm{IP}}$ proceeds through an intermediate antiferroelectric (AFE) phase (SG: Cmm2). While this AFE arrangement exhibits antiferromagnetism strictly within the 2D plane, the persistent OP asymmetry ensures the system's overall non-centrosymmetry, distinguishing it from hypothetical non-Janus V$_2$I$_2$O$_2$X$_2$ (X = Br or Cl) counterpart and maintaining AM, when the $z$ direction is considered periodically [Figure S7]. Therefore, the switchable $\mathbf{P}_{\mathrm{IP}}$ acts as an electrical control knob for the AM state, operating within a system possessing a constant symmetry-breaking bias from the fixed $\mathbf{P}_{\mathrm{OP}}$. This coexistence offers potential for designing multistate or multifunctional FEAM devices.

\textit{Electrical Control of Magnon Chirality}  --- 
\begin{figure}
\centering
\includegraphics[width=0.46\textwidth]{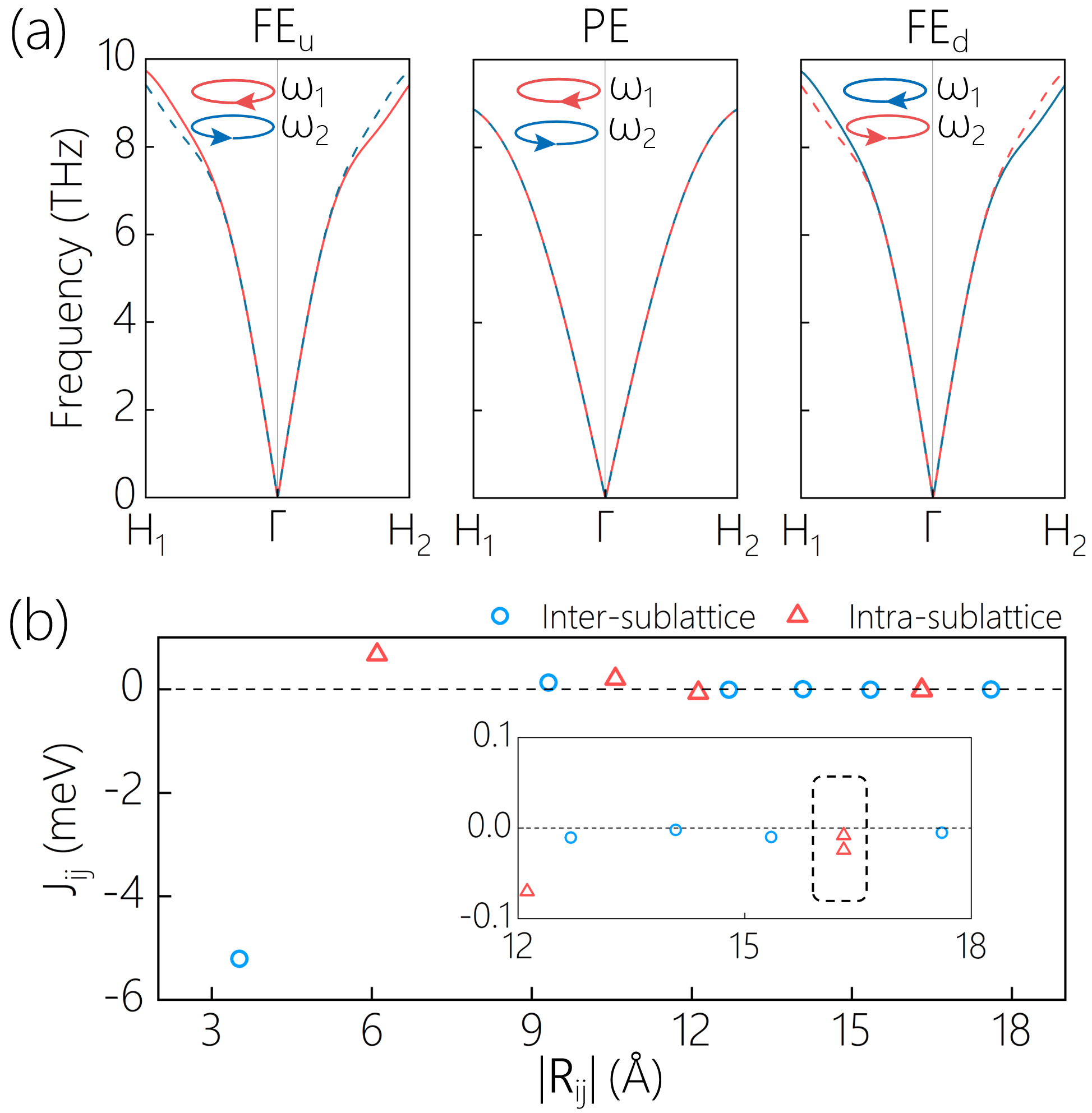}
\caption{The magnon properties of CrPS$_3$ monolayer. (a) Calculated magnon spectra concerning Figs. \ref{fig2} (a-c), where alternating splitting of two magnon modes with right-handed (counterclockwise, blue) and left-handed (clockwise, red) chirality and linear dispersion near the $\Gamma$ point in FE$_u$ and FE$_d$ phases. Right-handed and left-handed chirality are energy-degenerate modes for the PE phase in an antiferromagnetic state. (b) DFT Heisenberg exchange parameters as a function of the separation of Cr atoms $|\mathbf{R}_{ij}|$. $|\mathbf{R}_{ij}|$ is the distance of a paired Cr atoms. The exchange parameters highlighted by the dashed rectangle are responsible for the chirality-splitting magnon modes.
}\label{fig3}
\end{figure}
Magnetic excitations provide crucial insights into material properties and applications. Analogous to the electronic spin splitting in AMs, recent studies on 3D AMs like RuO$_2$ and MnTe~\cite{magnon_RuO2, INS-MnTe-magnon} have revealed chirality-split magnon bands, where modes with opposite chirality ($\hbar\omega_L(\mathbf{k}) \neq \hbar\omega_R(\mathbf{k})$) exhibit splitting that alternates across the BZ. This phenomenon necessitates anisotropic exchange interactions compatible with the specific spin group symmetries of the AM\cite{magnon_RuO2}. We thus hypothesize that in 2D FEAMs, the electrically switchable electronic spin splitting should be accompanied by \textit{switchable magnon chirality}. The underlying reason is that the FE switching exchanges the local crystal chiral symmetry, which in turn dictates the reversal of anisotropic Heisenberg exchange parameters. The combined $\mathcal{PT}$ operation relates the two states such that the effective Hamiltonian or state transforms as $\mathbf{P}$ $\to$ $-\mathbf{P}$ and $\mathbf{M}$ $\to$ $-\mathbf{M}$, where $\mathbf{M}$ corresponds to lifting degeneration of chiral magnon modes in one FE state. Consequently, reversing $\mathbf{P}$ is expected to interchange the left- and right-handed magnon branches, leading to $\Delta E^M_{FE_u}(\mathbf{q}) = - \Delta E^M_{FE_d}(\mathbf{q})$, where $\Delta E^M(\mathbf{q}) = \hbar\omega_{L}(\mathbf{q}) - \hbar\omega_{R}(\mathbf{q})$.

Magnon spectra for CrPS$_3$ and V$_2$I$_2$O$_2$BrCl were calculated using an effective Heisenberg model derived from DFT [Fig.~\ref{fig3} and Figure S8]. These calculations reveal significant chirality splitting $\Delta E^M(\mathbf{q})$, consistent with altermagnetic symmetry and driven by anisotropic exchange interactions extending to several neighbor shells at around 16.5 \AA[Figs.~\ref{fig3}(b)]. Confirming our hypothesis, the calculations show that reversing the polarization $\mathbf{P}$ induces a corresponding reversal in the chirality of the magnon modes. This electrically switchable magnon chirality, observed along paths like H1-$\Gamma$-H2, occurs synchronously with the switching of the electronic spin splitting [Figs.~\ref{fig2}(d-f)], same as that of RuO$_2$\cite{magnon_RuO2}, demonstrating the robust dual-switchable nature of these FEAMs.

The unique properties of altermagnetic magnons, combined with the demonstrated dual switchability in 2D FEAMs, offer exciting prospects. Magnons in AMs synergistically combine features of both FM and AFM: they can be chiral and carry net spin currents (like FMs), yet exhibit high propagation speeds and potentially linear dispersion over large parts of the BZ (like AFMs), often with strong anisotropy. Furthermore, their wavelengths are orders of magnitude shorter than photons of similar frequency\cite{magnon_RuO2}, enabling nanoscale wave-based computing architectures. The ability to electrically switch both the electronic spin polarization and the magnon chirality using the FE state in FEAMs adds a powerful control dimension. This dual control makes these materials highly promising platforms for advancing spintronics\cite{bai_altermagnetism_2024, AFMspintronics-2018, chen_emerging_2024}, spin caloritronics~\cite{SSE-SNE-magnon1}, magnonics\cite{chumak_magnon_2015, huang_antiferromagnetic_2024}, and potentially novel multifunctional memory and logic devices where information can be encoded and manipulated through coupled electronic and magnonic states.

\textit{Detecting the spin splitting and chiral magnon lifting}  ---
\begin{figure}
\centering
\includegraphics[width=0.46\textwidth]{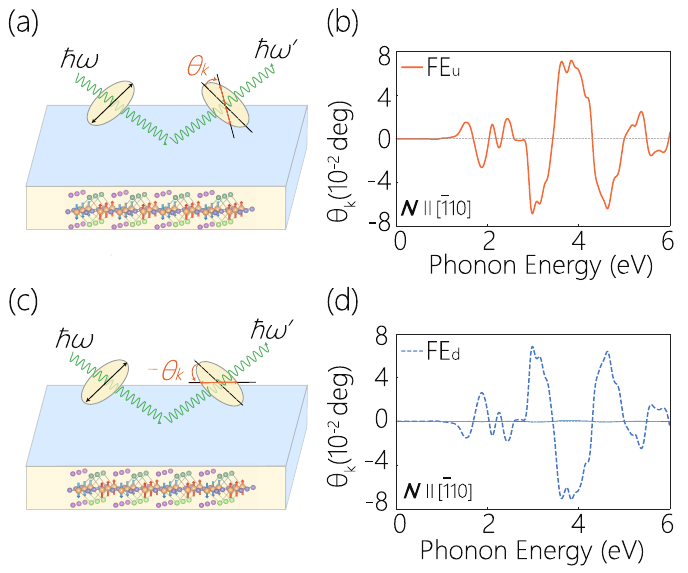}
\caption{Detection of the spin reversal in FEAMs.  (a) and (c) The schematic illustration of MOKE. Green screw arrows denote the propagation direction of light, and orange arrows marked with $\theta_K$ and $-\theta_K$ refer to the opposite direction of corresponding linear polarization direction. (b) and (d) The calculated MOKE signals characterized by Kerr angle $\theta_K$ and $-\theta_K$ for V$_2$I$_2$O$_2$BrCl. $N||$[$\overline{1}$10] is the N\'{e}el vector along the [$\overline{1}$10] direction for V$_2$I$_2$O$_2$BrCl monolayer.
}\label{fig4}
\end{figure}
%\subsection*{MOKE Signature of FEAM Switching}
The predicted dual-switchable FEAM state can be experimentally confirmed using optical techniques. Angle-resolved photoelectron spectroscopy (ARPES) can map the characteristic spin splitting $\Delta E^{S}(\mathbf{k})$. magnetic optical Kerr effect (MOKE) provides a sensitive probe of the magnetic state reversal linked to FE switching[Fig.~\ref{fig4}(a) and ~\ref{fig4}(c)], detecting changes in reflected light polarization due to broken $\mathcal{T}$ symmetry. Symmetry analysis dictates MOKE observability based on N\'{e}el vector orientation [Tables SII and SIII]; V$_2$I$_2$O$_2$BrCl allows detection broadly, while CrPS$_3$ is more restricted. As exemplified by V$_2$I$_2$O$_2$BrCl calculations [Fig.~\ref{fig4}(b) and (d)], a clear Kerr signal ($\theta_K$) is expected in FE$_u$ state. Upon switching the polarization, the altermagnetic state reverses, leading to a direct sign flip of the Kerr angle ($-\theta_K$). This electrically controlled reversal of the MOKE signal serves as a direct signature of the coupled magnetoelectric switching in the FEAM.

%\subsection*{Magnon-Based Detection Pathways}
For specific N\'{e}el vector orientations where MOKE detection might be precluded by symmetry in FEAMs, probing the coupled magnonic switching offers an alternative. The key signature is the electrically reversible magnon chirality splitting $\Delta E^M(\mathbf{k})$. Although direct mapping via inelastic neutron scattering (INS)~\cite{INS-MnTe-magnon} is challenging for 2D materials, thermal spin transport measurements like the spin Seebeck effect (SSE) can indirectly probe this reversal~\cite{SSE-SNE-magnon1, SSE-SNE-magnon2}. Reversing the polarization flips the sign of $\Delta E^M(\mathbf{k})$, altering magnon transport properties. This is expected to cause a sign change in the measured SSE voltage under a thermal gradient. Such a reversal in the SSE signal upon electrical switching would serve as compelling evidence for the ferroelectrically controlled magnonic state, complementing optical probes.

\textit{Conclusion} ---
We have proposed a scenario to achieve \textit{2D dual-switchable FEAMs}. Through systematic screening and first-principles calculations, we identified promising candidates like CrPS$_3$ and V$_2$I$_2$O$_2$BrCl monolayers. Our key finding is that the polarization acts as a master switch, intrinsically coupling to and \textit{simultaneously} controlling both the ground-state electronic properties (altermagnetic spin splitting, \textbf{S}) and the collective magnetic excitations (magnon chirality splitting, \textbf{M}). This unprecedented electrical control over both electron spin textures and magnon chirality stems from the ferroelectrically modulated crystal symmetry governing anisotropic exchange interactions. The predicted reversibility of both \textbf{S} and \textbf{M} upon polarization reversal \textbf{P}, verifiable via MOKE, ARPES, and potentially magnon-based probes (INS, SSE), establishes 2D FEAMs as a unique platform for exploring fundamental magnetoelectric phenomena involving coupled ground and excited states, opening new avenues in condensed matter physics and materials discovery.

This work was financially supported by the National Natural Science Foundation of China (Grant No. 12304165, 12304086), the Natural Science Foundation of Inner Mongolia Autonomous Region (Grants No. 2023QN01003), the "Grassland Talents" project of the Inner Mongolia autonomous region (Grant No. 21200-242920 ), the Startup Project of Inner Mongolia University (Grant No. 21200-5223733). 

%\bibliography{crpsref}% Produces the bibliography via BibTeX.

\begin{thebibliography}{100}%
%\bibliography{new}% Produces the bibliography via BibTeX.
\bibitem{multiferro-1994}H. Schmid, Multi-ferroic magnetoelectrics, Ferroelectrics \textbf{162}, (1994)

\bibitem{eerenstein_multiferroic_2006}W. Eerenstein, N. D. Mathur, and J. F. Scott, Multiferroic and magnetoelectric materials, Nature, \textbf{442}, 759 (2006).

\bibitem{multiferroic_2015}S. Dong, J. M. Liu, S. W. Cheong, and Z. Ren, Multiferroic materials and magnetoelectric physics: symmetry, entanglement, excitation, and topology, Adv. Phys. \textbf{64}, 519 (2015).

\bibitem{AM-PRX-2022}L. Šmejkal, J. Sinova, T. Jungwirth, Emerging Research Landscape of Altermagnetism,  Phys. Rev. X, \textbf{12}, 040501 (2022).

\bibitem{AM-PRX2-2022}L. Šmejkal, J. Sinova, T. Jungwirth, Beyond Conventional Ferromagnetism and Antiferromagnetism: A Phase with Nonrelativistic Spin and Crystal Rotation Symmetry, Phys. Rev. X, \textbf{12}, 031042 (2022).

\bibitem{krempasky_altermagnetic_2024}J. Krempaský, L. Šmejkal, S. W. D’Souza, M. Hajlaoui, G. Springholz, K. Uhlířová, F. Alarab,  P. C. Constantinou, V. Strocov, D. Usanov, W. Pudelko, R. González-Hernández, A. Birk Hellenes, Z. Jansa, H. Reichlová, Z. Šobáň, R. Gonzalez Betancourt, P. Wadley, J. Sinova, D. Kriegner, J. Minár, J. H. Dil, T. Jungwirth, Altermagnetic lifting of Kramers spin degeneracy, Nature. \textbf{626}, 517 (2024).

\bibitem{zhou_manipulation_2025}Z. Zhou, X. Cheng, M. Hu, R. Chu, H. Bai, L. Han,  J. Liu, F. Pan, C. Song, Manipulation of the altermagnetic order in CrSb via crystal symmetry, Nature, \textbf{638}, 645 (2025).

\bibitem{bai_altermagnetism_2024}L. Bai, W. Feng,  S. Liu, L. Šmejkal, Y. Mokrousov, Y. Yao,  Altermagnetism: Exploring New Frontiers in Magnetism and Spintronics,  Adv. Funct. Mater. \textbf{34}, 2409327 (2024).

\bibitem{liu_inverse_2024}Q. Liu, J. Kang,  P. Wang, W. Gao, Y. Qi, J. Zhao, and X. Jiang, Inverse Magnetocaloric Effect in Altermagnetic 2D Non-van der Waals FeX (X = S and Se) Semiconductors, Adv. Funct. Mater. \textbf{34}, 2402080 (2024).

\bibitem{zhang_simultaneous_2024}Y. Zhang, H. Bai, L. Han, C. Chen, Y. Zhou, C. H.Back, F. Pan, Y. Wang, and C. Song, Simultaneous High Charge-Spin Conversion Efficiency and Large Spin Diffusion Length in Altermagnetic RuO$_{2}$, Adv. Funct. Mater. \textbf{34}, 2313332 (2024).

\bibitem{guo_direct_2024}Y. Guo, J. Zhang, Z. Zhu, Y. Jiang,  L. Jiang, C. Wu, J. Dong, X. Xu, W. He, B. He, Z. Huang, L. Du, G. Zhang, K. Wu, X. Han, D. Shao, G. Yu, H. Wu, Direct and Inverse Spin Splitting Effects in Altermagnetic RuO$_{2}$, Adv. Sci. \textbf{11}, 2400967 (2024).

\bibitem{song_altermagnets_2025}C. Song, H. Bai, Z. Zhou, L. Han, H. Reichlova, J. H. Dil, J. Liu, X. Chen, F. Pan, Altermagnets as a new class of functional materials, Nat. Rev. Mater. 1 (2025).

\bibitem{jiang_metallic_2025}B. Jiang, M. Hu, J. Bai, Z. Song, C. Mu, G. Qu, W. Li, W. Zhu, H. Pi, Z. Wei, Y.-J. Sun, Y. Huang, X. Zheng, Y. Peng, L. He, S. Li, J. Luo, Z. Li, G. Chen, H. Li, H. Weng, T. Qian, A metallic room-temperature d-wave altermagnet, Nat. Phys. 1 (2025).

\bibitem{zhang_crystal-symmetry-paired_2025}F. Zhang, X. Cheng, Z. Yin, C. Liu, L. Deng, Y. Qiao, Z. Shi, S. Zhang, J. Lin, Z. Liu, M. Ye, Y. Huang, X. Meng, C. Zhang, T. Okuda, K. Shimada, S. Cui, Y. Zhao, G.-H. Cao, S. Qiao, J. Liu, and C. Chen, Crystal-symmetry-paired spin–valley locking in a layered room-temperature metallic altermagnet candidate, Nat. Phys. 1 (2025).

\bibitem{qian_fragile_2025}Z. Qian, Y. Yang, S. Liu, and C. Wu, Fragile Unconventional Magnetism in RuO$_{2}$ by Proximity to Landau-Pomeranchuk Instability, (2025), arXiv:2501.13616.

\bibitem{ABAM_PRL_RuO2}
 J. Liu, J. Zhan, T. Li, J. Liu, S. Cheng, Y. Shi, L. Deng, M. Zhang, C. Li, J. Ding, Q. Jiang, M. Ye, Z. Liu, Z. Jiang, S. Wang, Q. Li, Y. Xie, Y. Wang, S. Qiao, J. Wen, Y. Sun, and D. Shen, Absence of Altermagnetic Spin Splitting Character in Rutile Oxide ${\mathrm{RuO}}_{2}$, Phys. Rev. Lett. \textbf{133}, 176401 (2024).

\bibitem{bl_AM_PRL}
B. Pan, P. Zhou, P. Lyu, H. Xiao, X. Yang, and L. Sun, General Stacking Theory for Altermagnetism in Bilayer Systems, Phys. Rev. Lett. \textbf{133}, 166701 (2024).

\bibitem{tw_AM_PRL}
Y. Liu, J. Yu, and C.-C. Liu, Twisted Magnetic Van der Waals Bilayers: An Ideal Platform for Altermagnetism, Phys. Rev. Lett. \textbf{133}, 206702 (2024).



\bibitem{sun_TypeII-FEAM}W. Sun, C. Yang, W. Wang, Y. Liu, X. Wang, S. Huang, and Z. Cheng, Proposing Altermagnetic-Ferroelectric Type-III Multiferroics with Robust Magnetoelectric Coupling, Adv. Mater.  \textbf{n/a}, 2502575 (2025).

\bibitem{BR-popper_2024}F. Bernardini, M. Fiebig, and A. Cano, Ruddlesden-Popper and perovskite phases as a material platform for altermagnetism, (2025), ArXiv:2401.12910.

\bibitem{FESAM-Liu}M. Gu, Y. Liu, H. Zhu, K. Yananose, X. Chen, Y. Hu, A. Stroppa, and Q. Liu, Ferroelectric Switchable Altermagnetism, Phys. Rev. Lett. \textbf{134}, 106802 (2025).

\bibitem{Duan_AFMAM_2025}X. Duan, J. Zhang, Z. Zhu, Y. Liu, Z. Zhang, I. \v{Z}uti\'{c}, and T. Zhou, T. Antiferroelectric Altermagnets: Antiferroelectricity Alters Magnets, Phys. Rev. Lett. \textbf{134}, 106801 (2025).

\bibitem{zhu_FEAM_2025}Z. Zhu, X., Duan, J. Zhang, B. Hao, I. \v{Z}uti\'{c}, and T. Zhou, Two-Dimensional Ferroelectric Altermagnets: From Model to Material Realization, (2025), ArXiv:2504.06258.

\bibitem{wang_electric-field-induced_2025}D. Wang, H. Wang, L. Liu, J. Zhang, and H. Zhang, Electric-Field-Induced Switchable Two-Dimensional Altermagnets, Nano Lett. \textbf{25}, 498 (2025).

\bibitem{SSE-SNE-magnon1}Q. Cui, B. Zeng, P. Cui, T. Yu, and H. Yang, Efficient spin Seebeck and spin Nernst effects of magnons in altermagnets, Phys. Rev. B \textbf{108}, L180401 (2023).

\bibitem{SSE-SNE-magnon2}M. Weißenhofer, and A. Marmodoro, Atomistic spin dynamics simulations of magnonic spin Seebeck and spin Nernst effects in altermagnets,  Phys. Rev. B \textbf{110}, 094427 (2024).

\bibitem{magnon_RuO2}L. Šmejkal, A. Marmodoro, K.-H. Ahn, R. González-Hernández, I. Turek, S. Mankovsky, H. Ebert, S. W. D'Souza, Ond\v{r}ej \v{S}ipr j. Sinova, and T. c. v. Jungwirth, Chiral Magnons in Altermagnetic RuO$_{2}$, Phys. Rev. Lett. \textbf{131}, 256703 (2023).

\bibitem{sodequist_two-dimensional_2024}J. Sødequist, and T. Olsen, Two-dimensional altermagnets from high throughput computational screening: Symmetry requirements, chiral magnons, and spin-orbit effects, Appl. Phys. Lett. \textbf{124}, 182409 (2024).

\bibitem{wang_alternating_2024}Q. Wang, D.-W. Wu, G.-H Guo, M.-Q. Long, and Y.-P. Wang, Alternating spin splitting of electronic and magnon bands in two-dimensional altermagnetic materials, Chin. Phys. B \textbf{33}, 097507 (2024).

\bibitem{INS-MnTe-magnon}Z. Liu, M. Ozeki, S. Asai, S. Itoh, and T. Masuda, Chiral Split Magnon in Altermagnetic MnTe, Phys. Rev. Lett. \textbf{133}, 156702 (2024).

\bibitem{costa_giant_2024}A. T. Costa, J. C. G. Henriques, and J. Fernández-Rossier, Giant spatial anisotropy of magnon lifetime in altermagnets, (2024), ArXiv:2405.12896.

\bibitem{SSE-magnon}J. Li, Z. Shi, V. Ortiz, M. Aldosary, C. Chen, V. Aji, P. Wei, and J. Shi, Spin Seebeck Effect from Antiferromagnetic Magnons and Critical Spin Fluctuations in Epitaxial FeF$_{2}$ Films, Phys. Rev. Lett. \textbf{122}, 217204 (2019).

\bibitem{kimel_optical_2024}A. V. Kimel, T. Rasing, and B. A. Ivanov, Optical read-out and control of antiferromagnetic Néel vector in altermagnets and beyond, J. Magn. Magn. Mater. \textbf{598}, 172039 (2024).

\bibitem{gray_time-resolved_2024}I. Gray, Q. Deng, Q. Tian, M. Chilcote, J. S. Dodge, M. Brahlek, and L. Wu, Time-resolved magneto-optical effects in the altermagnet candidate MnTe, Appl. Phys. Lett. \textbf{125}, 212404 (2024).

\bibitem{hortensius_coherent_2021}J. R. Hortensius, D. Afanasiev, M. Matthiesen, R. V. Leenders, R. Citro, A. Kimel, R. Mikhaylovskiy, B. A. Ivanov, and A. D. Caviglia Coherent spin-wave transport in an antiferromagnet, Nat. Phys.  \textbf{17}, 1001-1006 (2021).

\bibitem{weber_all_2024}M. Weber, S. Wust, L. Haag, A. Akashdeep, K. Leckron, C. Schmitt, R. Ramos, T. Kikkawa, E. Saitoh, M. Kläui, L. Šmejkal, J. Sinova,  M. Aeschlimann, G. Jakob, B. Stadtmüller, and H. C. Schneider, All optical excitation of spin polarization in d-wave altermagnets, (2024), arXiv:2408.05187.

\bibitem{SM}
See Supplemental Material at http://link.aps.org/supplemental/ DOI for a detailed description of the computaional method used in this work; the screening process from C2DB; the electrinic structure, phonon, anisotropic Heisenberg exchange parameters,  and MOKE of FEAM candidates.


\bibitem{DFPT1}P. Giannozzi, S. de Gironcoli, P. Pavone, and S. Baroni, Ab initio calculation of phonon dispersions in semiconductors, Phys. Rev. B \textbf{43}, 7231 (1991).

\bibitem{DFPT2}X. Gonze, and C. Lee, Dynamical matrices, Born effective charges, dielectric permittivity tensors, and interatomic force constants from density-functional perturbation theory, Phys. Rev. B \textbf{55}, 10355 (1997).

\bibitem{PHONOPY}A. Togo, and I. Tanaka, First principles phonon calculations in materials science, Scripta Materialia. \textbf{108}, 1 (2015).

\bibitem{VASP}G. Kresse, and J. Furthmüller, Efficient iterative schemes for ab initio total-energy calculations using a plane-wave basis set,  Phys. Rev. B \textbf{54}, 11169, (1996).

\bibitem{PAW}P. A. Korzhavyi, I. A. Abrikosov, B. Johansson, A. V. Ruban, and H. L. Skriver, First-principles calculations of the vacancy formation energy in transition and noble metals, Phys. Rev. B \textbf{59}, 11693 (1999).

\bibitem{PBE}J. P. Perdew, K. Burke, and M. Ernzerhof, Generalized Gradient Approximation Made Simple, Phys. Rev. Lett. \textbf{77}, 3865, (1996).

\bibitem{VASPKIT}V. Wang, N. Xu, J. Liu, G. Tang, and W.-T. Geng, VASPKIT: A user-friendly interface facilitating high-throughput computing and analysis using VASP code, Comput. Phys. Commun. \textbf{267}, 108033 (2021).


\bibitem{Wannier90}A. A. Mostofi, J. R. Yates, Y.-S. Lee, I. Souza, D. Vanderbilt, and N. Marzari, wannier90: A tool for obtaining maximally-localised Wannier functions,Comput. Phys. Commun. \textbf{178}, 685 (2008).

\bibitem{TB2J}X. He, N. Helbig, M. J. Verstraete, and E. Bousquet, TB2J: A python package for computing magnetic interaction parameters, omput. Phys. Commun. \textbf{264}, 107938 (2021).

\bibitem{PASP}F. Lou, X. Y. Li, J. Y. Ji, H. Y. Yu, J. S. Feng, X. G. Gong, and H. J. Xiang, PASP: Property analysis and simulation package for materials, J. Chem. Phys. \textbf{154}, 114103 (2021).

\bibitem{Toth_2015}S. Toth, and B. Lake, Linear spin wave theory for single-Q incommensurate magnetic structures, J. Phys.: Condens. Mat., \textbf{27}, 166002 (2015).

\bibitem{COLPA1978327}J. Colpa, Diagonalization of the quadratic boson hamiltonian.,  Physica A \textbf{93}, 327 (1978).

\bibitem{DFTPU}S. Dudarev, G. Botton, S. Savrasov, C. Humphreys, and A. Sutton, Electron-energy-loss spectra and the structural stability of nickel oxide: An LSDA+U study, Phys. Rev. B \textbf{57}, 1505 (1998).

\bibitem{SSG1}Z. Xiao, J. Zhao, Y. Li, R. Shindou, and Z.-D. Song, Spin Space Groups: Full Classification and Applications, Phys. Rev. X \textbf{14}, 031037 (2024).

\bibitem{SSG2}X. Chen, J. Ren, Y. Zhu, Y. Yu, A. Zhang, P. Liu,  J. Li, Y. Liu, C. Li, and Q. Liu, Enumeration and Representation Theory of Spin Space Groups, Phys. Rev. X \textbf{14}, 031038 (2024).

\bibitem{SSG3}Y. Jiang, Z. Song, T. Zhu, Z. Fang, H. Weng, Z.-X. Liu, J. Yang, and  C. Fang, Enumeration of Spin-Space Groups: Toward a Complete Description of Symmetries of Magnetic Orders, Phys. Rev. X \textbf{14}, 031039 (2024).

\bibitem{C2DB}M. N. Gjerding, A. Taghizadeh, A. Rasmussen, S. Ali, F. Bertoldo, T. Deilmann, N. R. Knøsgaard, M. Kruse, A. H. Larsen, S. Manti, T. G. Pedersen, U. Petralanda, T. Skovhus, M. K. Svendsen, J. J. Mortensen, T. Olsen, and K. S. Thygesen, Recent progress of the Computational 2D Materials Database (C2DB), 2D Materials, \textbf{8}, 044002 (2021).

\bibitem{PSEUDO}
C. Capillas, E.S. Tasci, G. de la Flor, D. Orobengoa, J.M. Perez-Mato and M.I. Aroyo. A new computer tool at the Bilbao Crystallographic Server to detect and characterize pseudosymmetry. Z. Krist. \textbf{226}, 186-196 (2011).  

\bibitem{CINEB}G. Henkelman, B. P. Uberuaga, and H. Jónsson, A climbing image nudged elastic band method for finding saddle points and minimum energy paths, J. Chem. Phys., \textbf{113}, 9901, (2000).

\bibitem{Ni2021_MnPSe3}Z. Ni, A. V. Haglund, H. Wang, B. Xu, C. Bernhard, D. G. Mandrus, X. Qian, E. J. Mele, C. L. Kane, and L. Wu, Imaging the Néel vector switching in the monolayer antiferromagnet MnPSe$_{3}$ with strain-controlled Ising order, Nat. Nanotechnol., \textbf{16}, 782, (2021).

\bibitem{AFMspintronics-2018}V. Baltz, A. Manchon, M. Tsoi, T. Moriyama, T. Ono, and Y. Tserkovnyak,Antiferromagnetic spintronics, Rev. Mod. Phys., \textbf{90}, 015005 (2018).

\bibitem{chen_emerging_2024}H. Chen, L. Liu, X. Zhou, Z. Meng, X. Wang, Z. Duan, G. Zhao, H. Yan, P. Qin, and Z. Liu, Emerging Antiferromagnets for Spintronics, Adv. Mater., \textbf{36}, 2310379 (2024).

\bibitem{chumak_magnon_2015}A. V. Chumak, V. I. Vasyuchka, A. A. Serga, and B. Hillebrands, Magnon spintronics, Nat. Phys., \textbf{11}, 453-461 (2015).


\bibitem{huang_antiferromagnetic_2024}L. Huang, L. Liao, H. Qiu, X. Chen, H. Bai, L, Han, Y. Zhou, Y. Su, Z. Zhou, F. Pan, B. Jin, and C. Song, Antiferromagnetic magnonic charge current generation via ultrafast optical excitation, Nat. Commun., \textbf{15}, 4270 (2024).
\end{thebibliography}

%\begin{thebibliography}{70}%

%F. Schindler, M. Brzezi\'{n}ska, W. A. Benalcazar, M. Iraola, A. Bouhon, %S. S. Tsirkin, M. G. Vergniory, and T. Neupert,
%Phys. Rev. Research \textbf{1}, 033074 (2019).
%\end{thebibliography}

\end{document}